\documentclass[11pt]{elsarticle}
\usepackage[top=1in, bottom=1in, left=1in, right=1in]{geometry}  

\usepackage{amsmath}
\usepackage{amssymb}

\usepackage{graphicx, url}

\def\e#1{\times 10^{#1}}
\def\bbP{\mathbb{P}}

\journal{arXiv}

\begin{document}

\begin{frontmatter}

\title{The influence of relatives on the efficiency and error rate of familial searching}

\author[IB]{Rori V. Rohlfs\corref{RVRMS}}
\ead{rrohlfs@berkeley.edu}
\author[NYU]{Erin Murphy} 
\ead{eem9@nyu.edu}
\author[CS,STAT]{Yun S. Song}
\ead{yss@eecs.berkeley.edu}
\author[IB]{Montgomery Slatkin}
\ead{slatkin@berkeley.edu}   

\cortext[RVRMS]{Corresponding author, rrohlfs@berkeley.edu, +1 (510) 643-0060 }

\address[IB]{Department of Integrative Biology, University of California, Berkeley, CA 94720, USA}
\address[NYU]{School of Law, New York University, New York, NY 10012, USA}
\address[CS]{Computer Science Division, University of California, Berkeley, CA 94720, USA}
\address[STAT]{Department of Statistics, University of California, Berkeley, CA 94720, USA}

\pagebreak

\title{The influence of relatives on the efficiency and error rate of familial searching}

\begin{abstract}
We investigate the consequences of adopting the criteria used by the state of California, as described by Myers {\it et al.} (2011), for conducting familial searches. We carried out a simulation study of randomly generated profiles of related and unrelated individuals with 13-locus CODIS genotypes and YFiler$^{\textregistered}$ Y-chromosome haplotypes, on which the Myers protocol for relative identification was carried out.  For Y-chromosome sharing first degree relatives, the Myers protocol has a high probability ($80\sim 99\%$) of identifying their relationship.  For unrelated individuals, there is a low probability that an unrelated person in the database will be identified as a first-degree relative.  For more distant Y-haplotype sharing relatives (half-siblings, first cousins, half-first cousins or second cousins) there is a substantial probability that the more distant relative will be incorrectly identified as a first-degree relative.  For example, there is a $3\sim 18\%$ probability that a first cousin will be identified as a full sibling, with the probability depending on the population background.  Although the California familial search policy is likely to identify a first degree relative if his profile is in the database, and it poses little risk of falsely identifying an unrelated individual in a database as a first-degree relative, there is a substantial risk of falsely identifying a more distant Y-haplotype sharing relative in the database as a first-degree relative, with the consequence that their immediate family may become the target for further investigation. This risk falls disproportionately on those ethnic groups that are currently overrepresented in state and federal databases.  
\end{abstract}


\end{frontmatter}

\section*{Introduction}
 DNA databases have unquestionably assumed  a vital role in the American criminal  justice  system.   Genetic  evidence  has  served  to bolster  the evidence in existing cases and to identify suspects through ``cold hit'' database matches \cite{Bieber2006}.  Typically,  investigative queries in DNA databases have been limited  to searches  intended  to find the source of the crime-scene  sample. However, increasing attention has been given to the question of whether law enforcement should  also be able to search  for partial matches, that is, DNA searches intended to find the source by identifying a relative in the  database \cite{Bieber2006,Greely2006}.  Such searches  are  commonly  called ``kinship''  or ``familial''  searches.

 The concept of familial searching is not particularly new. In fact, familial searches fueled some of the earliest illustrations of the investigative power of DNA typing \cite{Curran2008}. In 2002, investigators in the United Kingdom  identified a serial rapist in part through a database search  that led them to the perpetrator via the DNA profile of his son \cite{Williams2005}.  In another  widely cited  case, UK investigators recovered  DNA from a brick thrown off an overpass  that landed on a truck, leading to the driver's fatal heart attack, and found the source through a database search that located a relative \cite{Murphy2011}.   More recently, California authorities used a familial search to identify the putative son of a serial killer nicknamed the ``Grim Sleeper,'' and arrested the suspect after a sting operation in which police collected a discarded pizza crust \cite{Murphy2011}. 

 Familial searches by no means dominate the use patterns of DNA databases, in part because they are difficult to conduct, require access to a sizeable database, are subject to various clear or unclear legal restrictions, and raise ethical concerns \cite{Gershaw2011}.  Because familial searches are by design inexact, the most effective methods typically employ several steps beyond a simple database search.  To find a lead, one commonly used approach, which may entail additional rounds of testing, relies on examining the Y haplotype of all significant partial matches. Even if a partial match is identified, law enforcement still must investigate the relatives of that person to determine if any are likely to be the crime scene sample source.  

  Nevertheless, familial searches are presently conducted by a number of jurisdictions and continue to garner interest.  The UK is the most prominent and longstanding advocate of the technique: from 2004 to 2011, 179 cases were submitted for search \cite{ref7}.  As of 2009, New Zealand had conducted 12 familial searches in serious cases \cite{X}.  The Netherlands recently passed legislation authorizing familial searches \cite{ref6a,ref6b}.  Japan, Australia and Canada have robust DNA collection programs, but  only Canada  has explicitly rejected familial searches on what appear to be privacy grounds \cite{ref9}.  

 With regard to Europe, it is worth noting that adoption may be slowed by the December 2008 judgment by the European Court of Human Rights that declared  the retention of DNA profiles and samples from unconvicted persons a violation of Article 8 of the Convention for the Protection of Human Rights and Fundamental Freedoms.  That opinion, (S. \& Marper v. United  Kingdom), invoked the provision of the Convention that safeguards its members' ``private life'' \cite{ref11}.  Although the influence of the ruling beyond its immediate holding is unclear, the Marper decision does represent  the first substantial curtailment of DNA expansion programs by a legal entity.  Moreover, Marper  may be used by privacy advocates and opponents to widespread  DNA typing to bolster their legal claims to circumscribe  such programs. 

 In the United  States,  the push  to expand  DNA testing  has intensified. Originally, United States national database administrators prohibited the disclosure of identifying information for partial matches made across state lines \cite{Murphy2011}.  As a result, although many states either legally authorized or informally permitted partial match (``moderate stringency'') reporting and/or familial searches \cite{Ram2011}, investigators could not obtain informational leads on profiles generated  out of state.  In 2006, however, the FBI modified its policy and now permits interstate sharing \cite{CODIS2006}. As of May 2012, a bill was pending before Congress that would allow the FBI to conduct familial searches in federal and state investigations \cite{D}.

 Because the rules governing familial search methods in the United States consist of a patchwork of state law, state and local regulation, and even internal laboratory  policies \cite{Murphy2011, Gershaw2011}, it is impossible to relate a precise legal picture.  In June of 2013, the U.S. Supreme Court in Maryland v. King \cite{MDKing2013} upheld DNA collection from arrestees for serious offenses.  Although the Court noted that Maryland forbids familial searches, that observation did not seem central to its holding, and no lower courts have ruled on the issue.  Assessment is further complicated by the slim line that differentiates unintentional and intentional partial match searches, because some jurisdictions allow the former but not the latter.  Nevertheless, some clarity is possible. 

 At the state level, both Maryland and Washington, D.C. have laws expressly forbidding familial searching \cite{A,B}, although the language of both statutes could be interpreted to permit reporting of unintentional partial matches. As a matter of either written or unwritten policy, roughly nine states expressly forbid both partial matching and familial searching: Alaska, Nevada, Utah, New Mexico, Michigan, Vermont, Massachusetts, and Georgia \cite{CRG}.  At least another seven states prohibit familial searches, but allow reporting of inadvertent partial matches \cite{CRG}. Fifteen states allow both forms of partial matching, although all of them rely upon formal or informal policies rather than express statutory authorization \cite{CRG}.  

 The states most actively pursuing familial matches are California, Colorado, Virginia, and Texas; Pennsylvania, Minnesota, and Tennessee are considering legislation.  In April 2008, California became the first state to formally  endorse  and  adopt explicit  rules for conducting  intentional  familial searches \cite{CADOJ2008}.

 The burgeoning interest in familial searching has reignited a national conversation about the propriety of the method that focuses on legal and ethical issues \cite{Rosen2009, Nakashima2008}.  The major concerns are two-fold: first, is familial searching actually efficacious, and second, does it adequately respect privacy and equality interests?  

 With regard to efficacy, the challenges of familial searching are reflected in its reported success rates, albeit based on limited data. The UK reports the greatest effectiveness with a $11\sim 27\%$ success rate) \cite{Butler2012}. California has conducted 29 searches, with 2 reported successes ($\sim 7\%$ success rate) \cite{E}. 

 With regard to the ethical issues, familial searches raise privacy, equality, and democratic accountability concerns \cite{Murphy2011, Hicks2010, Greely2006, Bieber2006, Gershaw2011, Garrison2013}.  In the United States, the most common critique is that the method is likely to have a discriminatory effect because DNA databases contain the profiles of certain racial minorities in disproportion to their presence in the population.  To date there have been only a handful of efforts to quantify the impact of familial searches, and all have  been undertaken  without reference to a specific search policy \cite{Bieber2006,Greely2006,Curran2008}.  Only one study, by a multidisciplinary team of researchers, attempted to calculate the general discriminatory impact and concluded  that roughly ``four times as much of the African-American population as the U.S. Caucasian  population would be `under surveillance' as a result of family forensic DNA'' \cite{Greely2006}.  It is this estimation that scholars,  policymakers  and  the popular  press  have  latched  upon  as  a  means  of quantifying  the racial impact of familial searching \cite{Rosen2009,Reid2008,Grimm2007,CBS2007}, and while helpful, it is nonetheless an approximation reached  before any specific policy was in place to be examined.

 Answering the efficacy and ethical concerns raised by familial search methods in part requires addressing complex statistical questions. The articulation of the first formal familial search  policy by California \cite{Myers2011}, an American state with the world's fourth largest DNA database (nearly 2 million profiles) and a large and diverse general population \cite{Steinberger2009}, affords an opportunity  to gain  valuable  insight  into the question of whether and under what circumstances familial searching should be allowed.  The racial and ethnic diversity of the California database roughly mirrors the racial and ethnic diversity of the United States national database \cite{West2008,Budowle1998}. Moreover, as a bellwether of criminal justice policy, California has already wielded influence both nationally and internationally as other jurisdictions contemplate various approaches.

\section*{Methods}
Here we implement the Myers {\it et al.} familial identification procedure used for familial searching in California \cite{Myers2011} to estimate power and false positive rate in addition to estimating the rates of misidentification of distant relatives as first-degree relatives.  As detailed more below, in the Myers {\it et al.} method, both parent-offspring and sibling relationships are considered by first calculating each likelihood ratio using autosomal data between the unknown sample and each entry in the state database.  Of these, the database samples with the highest likelihood ratios are considered in a secondary likelihood ratio analysis using Y-chromosome haplotypes.  The cumulative likelihood ratios are calculated under three population genetic assumptions and if they pass particular thresholds, the individual is considered a suspect.  This method is detailed in our descriptions below.  


\subsection*{Allele frequency data}

\subsubsection*{Autosomal data}
 In this study, we use allele frequency estimates to investigate identification procedures that are contingent on racially defined population sample allele frequency calculations.  For the autosomal STR allele frequencies, we rely upon estimates from a published survey of five population samples consisting of 182-213 individuals each and classified according to socially-identified race \cite{Budowle1998}.  The groups are described in the study as `Vietnamese,' `African American,' `Caucasian,' `Hispanic,' and `Navajo.'  Any labeling scheme introduces questions and classifies groups in different ways not independent of the social construction of these groups.  In this study we use the labels Vietnamese American, African American, European American, Latino American, and Native American.  

 The consent and population grouping procedures used to obtain these data are not clear. Since these data were collected, the customary ethical standards regarding informed consent processes have changed considerably, driven by several cases of severe misuse of samples provided by Indigenous communities \cite{Dalton2002, Wiwchar2004, Mello2010, ANDES2011, Arbour2006, Goering2008, Anderson2009, Kaye2009, McInnes2011}.  We use these data because of their public availability and utility to investigate error rates and efficacy in familial searching.  We look forward to working with data collected using transparent informed consent methodology.  

 The California state database consists of some entries with the 13 core CODIS loci and some with 15 loci \cite{Myers2011}.  To maintain manageable complexity, in this study we only consider the core 13 loci.  Similar analyses can be performed with 15 locus profiles.  

\subsubsection*{Y-chromosome data}
 For the Y-haplotypes, we consider data released by ABI consisting of YFiler$^{\textregistered}$ haplotypes genotyped in individuals grouped according to social labels `Vietnamese,' `African American,' `Caucasian,' `Hispanic,' and `Native American', with sample sizes of 103, 1918, 4102, 1594, and 105 individuals, respectively (Applied Biosystems$^{\textregistered}$, Foster City, CA) \cite{ABI}.  Again, we refer to these groups as Vietnamese American, African American, European American, Latino American, and Native American.  

 Individuals were genotyped and categorized into population labeling schemes differently for the autosomal and Y-chromosome markers.  In this study, we use samples with the same labels in both the autosomal and Y chromosome data to get our combined population sample allele frequencies for the Vietnamese American, African American, European American, and Latino American groups.  Accordingly, the group we call Native American is created from `Navajo' autosomal marker allele frequencies and `Native American' Y-chromosome allele frequencies.  This inconsistency brings to question the relevance of these results for highly specified populations.  However, this degree of inconsistency in population labeling is not remarkable when considering the wide variation typical to categorizing population groups (social identity-based labels like `Hispanic', `African American,' or `Caucasian').  The results of the analysis of these data should be confirmed and augmented by similar analyses of more transparent data.  

\subsection*{Simulation scheme}

\subsubsection*{Simulating relatives}
 To investigate the power and false positive rate of relative identification procedures, pairs of related individuals were simulated.  Specifically, 100,000 pairs of parent-offsprings, siblings, half-siblings, cousins, half-cousins (individuals sharing a single grandparent), and second cousins (individuals sharing a set of great-grand parents) were simulated using allele frequency distributions for each of the five populations described above.  The relative pairs were simulated to share a Y-haplotype by descent, and we refer to this sort of relationship as Y-sharing.  The autosomal markers for all of the individual pairs were simulated with a population background relatedness parameter $\theta=.01$, in accordance with the lower recommended correction in identification likelihood ratio estimations \cite{NRC1996}.  

\subsubsection*{Simulating unrelated individuals}

 Since unrelated individuals very rarely share enough alleles to resemble genetic relatives, more simulations are needed to accurately estimate the rates of positive relative identification between unrelated individuals.  To this end, 200,000,000 pairs of unrelated individuals were simulated based on allele frequencies from each pair of population samples.  

 Because of the immense polymorphism of Y-chromosome haplotypes, accurate estimates of background Y-chromosome relatedness ($\hat{\theta}_{Y}$) require greater sample sizes.  To simulate Y-haplotypes of unrelated individuals with realistic levels of background relatedness, haplotypes were independently drawn from the data.  This way, rates of coincidentally shared Y-haplotypes correspond with those observed in the available data.  Note that simulated rates of coincidentally shared Y-chromosome haplotypes are greatly influenced by the available data, which for some population samples is based on small numbers of individuals.  

\subsection*{Relative identification procedure}
 Parent-offspring and sibling identification protocols were followed with the method implemented in California which incorporates autosomal and Y-chromosome haplotype data \cite{Myers2011}.  These calculations were performed on pairs of individuals simulated with different genetic relationships, using the allele frequencies from each population sample.  

\subsubsection*{Autosomal likelihood ratio}
 Using autosomal data, the standard likelihood ratio (LR) comparing the probabilities of the observed genotypes ($G$) assuming a particular genetic relationship (parent-offspring or sibling) and assuming the individuals are unrelated is defined as \cite{Bieber2006, Buckleton2005}
\begin{equation}
  \widehat{LR}_{A} = \frac{\bbP(G | k_0, k_1, k_2)}{\bbP(G | k_0=1, k_1=0, k_2=0)},
\label{eq:LR_X}
\end{equation}
where $k_0$, $k_1$, and $k_2$ are parameters describing the probabilities that individuals with the specified relationship share 0, 1, or 2 alleles identical by descent (IBD) \cite{Weir2006}.  As specified by Myers {\it et al.}, this LR is estimated under three conditions using allele frequency distributions from African American, European American, and Latino American population samples with no $\theta$-correction for population substructure, as practiced in California \cite{Myers2011}.  

\subsubsection*{Y-haplotype likelihood ratio}
Ignoring mutation, the probability that two Y-sharing relatives have the same haplotype of population frequency $p$ is $p$.  On the other hand, the probability that two unrelated male individuals each have that same haplotype is $p^2$.  So, the Y-haplotype likelihood ratio $LR_Y$ is $1/p$.  In the Myers {\it et al.} procedure \cite{Myers2011}, $LR_Y$ was estimated as the inverse of the upper $95\%$ confidence limit of the haplotype frequency, obtained using the data pooled across populations excluding the sampled haplotype \cite{SWGDAM2009b, Myers2011}.  Specifically, after exclusion, if the Y-haplotype is observed with sample frequency $\hat{p}$ in the database, 
\begin{equation}
   \widehat{LR}_{Y} = \left[ \hat{p} + 1.96 \sqrt{\frac{\hat{p}(1-\hat{p})}{n}} \right] ^{-1},
\label{eq:LR_Y1}
\end{equation}
whereas if the Y-haplotype is not observed in the database, 
\begin{equation}
  \widehat{LR}_{Y} = \left(1-0.05^{1/n} \right) ^{-1},
\label{eq:LR_Y2}
\end{equation}
where $n$ denotes the total number of Y-haplotypes in the database. 

\subsubsection*{Combined result}
The combined test statistic defined by Myers {\it et al.} \cite{Myers2011} is the product of the autosomal marker and Y-haplotype LR estimates, divided by the database size ($N$):
\begin{equation}
   X = \frac{\widehat{LR}_{A} \cdot \widehat{LR}_{Y}}{N}.
\label{eq:X}
\end{equation}
$X$ is calculated for each of the three population samples described above.  In this study we consider a database of size $N=1,824,085$, the size of the California state database as of January 2012 \cite{FBI2012}.  

 An investigative positive identification (called simply a positive identification here) is called when $X$ is greater than 0.1 under all three assumed population samples, and greater than 1.0 for at least one population sample \cite{Myers2011}.

\section*{Results}
\subsection*{False positive rates of relative identification}
 Unrelated individuals were simulated based on allele frequency data from five population samples to investigate false positive rates of parent-offspring and sibling identification.  Autosomal and Y-chromosome LRs were estimated using \eqref{eq:LR_X}-\eqref{eq:LR_Y2}, and the combined test statistic $X$ defined in \eqref{eq:X} was calculated for unrelated pairs of individuals simulated from all pairs of population samples.  Using the procedure described by Myers et al. \cite{Myers2011}, false positive rates were estimated for parent-offspring (Table~\ref{tab:unrelpoFPRs}) and sibling (Table~\ref{tab:unrelsibFPRs}) identifications.  

 Even though false positive rates are low, on the order of $1\e{-5}$ to $1\e{-9}$, across population sample pairs, there is some variation (Tables~\ref{tab:unrelpoFPRs} and \ref{tab:unrelsibFPRs}).  In particular, the false positive rates for unrelated pairs of individuals simulated with Vietnamese American and with Native American allele frequencies are relatively high and low, respectively (Tables~\ref{tab:unrelpoFPRs} and \ref{tab:unrelsibFPRs}).  In sibling identification, the Vietnamese American sample shows a comparatively high false positive rate of $1.1\times 10^{-5}$, while no false positives are observed in the Native American sample (Table~\ref{tab:unrelsibFPRs}).  This can be explained by the particular Y-haplotype patterns considered for these population samples.  False positive identifications were observed only when unrelated individuals coincidentally share a Y-haplotype.  In the available Vietnamese American population sample of Y-haplotypes ($n=103$), several pairs of individuals share Y-haplotypes, while in the Native American population sample ($n=105$), no individuals share Y-haplotypes.  In the other population samples, Y-haplotypes are shared at frequencies intermediate to those in the Vietnamese American and Native American population samples.  Given the small sizes for these population samples, it is not clear if varying rates of coincidental Y-haplotype sharing are due to population genetic differences, or stochasticity of small samples.  

 To examine the validity of the total lack of observed false positive relative identifications for unrelated individuals simulated from the Native American population sample, we consider the possibility of observing complete Y-haplotype diversity (as observed) by chance.  Using simulations, 100,000 subsamples of 105 (the Native American sample size) Y-haplotypes were randomly chosen from the larger African American, European American, and Latino American samples.  Of the subsamples, 0.67, 0.57, 0.37 of the African American, European American, and Latino American samples, respectively, consisted of all unique haplotypes, as observed in the Native American sample.  This indicates the plausibility that a small sample from a group with the intermediate degree of Y-haplotype diversity observed in these larger population samples could all have unique Y-haplotypes by chance.  Larger Y-haplotype samples are required to confidently estimate false positive rates between unrelated individuals across population samples.

\subsubsection*{False positives in the database context}
 Our results agree with previous work, showing that with the prescribed methodology, false positive rates of parent-offspring and sibling identification are low, on the order of $1\e{-5}$ to $1\e{-9}$ (Tables~\ref{tab:unrelpoFPRs} and \ref{tab:unrelsibFPRs}) \cite{Myers2011}.  But even with these low false positive rates, differences were observed between population samples, raising the question of how these differences in false positive rates interact with distortions in DNA database representation.  

 To investigate this question, California census and prison population proportions of Asian, African American, European American, Latino American, and Native American individuals were normalized to fit the assumption that all individuals are described by exactly one of these categories (Table~S1 in File S1) \cite{CADOC2008,USCensus2010}.  In combining census and population genetic data, groups labeled as `Vietnamese' and `Asian' were equated to each other.  Clearly, these simplifications limit the applicability of the population sample-specificity of this analysis, however it provides a first approach.  

 Using each of the census and prison demographics, the proportion of false positive parent-offspring and sibling identifications that involve at least one member of each population group were estimated (Tables~S2 and S3 in File S1).  As expected, in the demographic context of a prison system in which African Americans are drastically over-represented (Table~S1 in File S1, exact binomial test $p<2.2\e{-16}$), the rates of false identification of individuals in this groups is much higher, roughly two orders of magnitude higher (Tables~S2 and S3 in File S1).  Nevertheless, the overall rate of false identification of unrelated individuals remains low.  

\subsection*{Spurious identification of distant relatives}
 The simulations of unrelated individuals showed low false positive rates of parent-offspring and sibling identification.  However, distant Y-sharing relatives may be more often mistaken for parent-offsprings or siblings.  To investigate this, individuals with various Y-sharing relationships (parent-offspring, siblings, half-siblings, cousins, half-cousins, and second cousins) from population sample backgrounds were simulated and used in the same relative identification procedure.  Note that when considering Y-sharing relatives, the $\widehat{LR}_Y$ calculation is greatly influenced by the database size, as opposed to the Y-haplotype reference frequency.  

 The observed distributions of the test statistic $X$ for second-degree and distant relatives is shifted left of those for first-degree relatives, but still has significant mass greater than $1$ (Figure~\ref{fig:LRdists}).  So as relatedness decreases, the $X$ more effectively distinguishes first-degree from distant Y-sharing relatives.  Concordant with a previous study \cite{Rohlfs2012}, distinguishability is also higher with appropriately-specified allele frequencies in population samples with higher polymorphism at the markers considered.  By considering these distributions, it is clear that regardless of the exact decision procedure, distant Y-sharing relatives show elevated $X$ values.  

 Positive rates vary across true relationships, population samples, and tests of parent-offspring versus sibling relationships (Figure~\ref{fig:IDrates}, Tables~\ref{tab:poIDrates} and \ref{tab:sibIDrates}).  The power of the parent-offspring test varies from 0.94 to 0.99 and the sibling test varies from 0.68 to 0.85 for various population samples.  Of course, a different threshold procedure could raise the power of these tests, but will simultaneously raise the false positive rates.  Regardless of the particular threshold procedure, the relative trends observed across true relationships, population samples, and tests of parent-offspring versus sibling relationships will hold for LR-based methods.  

 When implementing the full Myers {\it et al.} procedure to call putative relatives, Y-sharing relatives  are frequently mistakenly identified as parent-offsprings or siblings (Table~\ref{tab:sibIDrates}).  Second degree Y-sharing relatives like half-siblings are called as siblings in $5\sim 24\%$ of simulations (Table~\ref{tab:sibIDrates}).  The frequency of relative identification decreases with the degree of relatedness (or equivalently, those with higher kinship coefficients), but even Y-sharing half-cousins are called as siblings in $1\sim 10\%$ of simulations, depending on the population sample (Table~\ref{tab:sibIDrates}).  

 Positive identification between distant Y-sharing relatives occurs more often when considering sibling relationships rather than parent-offspring because of the less stringent allele sharing requirements.  For example, Y-sharing half-siblings are called as siblings in $5\sim 24\%$ of simulations and called as parent-offspring in $4\sim 10\%$ of simulations (Tables~\ref{tab:poIDrates} and \ref{tab:sibIDrates}).  Sharing at least one allele at each locus, as required for parent-offspring relationships, is less likely by chance than sharing on average one allele at each locus, as expected for sibling relationships.  

 Higher rates of positive identification are observed for individuals simulated with Native American or Vietnamese American allele frequencies (Figure~\ref{fig:IDrates}, Tables~\ref{tab:poIDrates} and \ref{tab:sibIDrates}).  This is likely due to allele frequency misspecification inherent in the method, which calculates the test statistic $X$ under African American, European American, and Latino American allele frequencies only, and due to varying population sample gene diversity, as found in a study of autosomal loci \cite{Rohlfs2012}.  For relatives simulated from African American, European American, or Latino American population samples, the method correctly specifies their allele frequencies, so they show comparatively lower identification rates (Figures~\ref{fig:IDrates} and \ref{fig:LRdists}, Tables~\ref{tab:poIDrates} and \ref{tab:sibIDrates}).  

 To show that these differences in identification rates across population samples are not driven by differing sample sizes, the same rates were estimated with a reduced Y-haplotype reference of 103 haplotypes per population sample.  Again, we see the same trends across population samples, confirming that they are not caused by varying reference Y-haplotype sample sizes (Tables~S4 and S5 in File S1).  Note that the absolute false identification rates differ in the full and subsample analysis because the estimated Y-haplotype frequency a function of the pooled sample size.  

\section*{Discussion}
 We have investigated by computer simulation the consequences of using a familial search policy similar to that described by Myers {\it et al.} \cite{Myers2011}, which is the policy currently used by the state of California for conducting familial searches. Our simulations assumed that allele frequencies at the 13 CODIS loci and the Y haplotypes for five ethnic groups are as given in Budowle {\it et al.} and the ABI reference database \cite{Budowle1998, ABI}. We reach three main conclusions. First, if the profile of a first-degree relative of a randomly generated profile is in the database searched, there is a relatively high probability of identifying the relative as such. Thus we agree with Myers {\it et al.} \cite{Myers2011},  Bieber {\it et al.} \cite{Bieber2006}, and Curran and Buckleton \cite{Curran2008} that familial searching can be an effective way to identify first-degree Y-sharing relatives of an individual who left a crime scene sample.  However, note that the simulation study of Bieber {\it et al.} \cite{Bieber2006} suggests higher identification efficiency than observed in an empirical study by Curran and Buckleton \cite{Curran2008}, possibly due to population structure in the empirical dataset \cite{Rohlfs2012}.  Slooten and Meester \cite{Slooten2012} have also shown that there may be high variability in power to identify relatives when considering profiles of varying rareness in specific databases.  

 Second, we found that the probability of identifying an unrelated Y-chromosome-carrying individual as a first-degree relative is quite low, agreeing with the results of Myers et al. \cite{Myers2011}. However, our ability to obtain precise estimates of this probability for different ethnic groups is limited by the relatively small samples sizes available to estimate Y haplotype frequencies, especially for the Vietnamese American and Native American samples.  For population samples other than those, the probabilities are so low that we could reasonably expect at most one unrelated individual would be incorrectly identified as a first-degree relative even in a database as large as California's, which is approaching 2 million profiles.  The high false positive rate in the Vietnamese American population sample is subject to sampling error with the relatively low number of Y-haplotypes for this group (103 haplotypes), so we hesitate to put great confidence behind that particular rate.  

 Our third conclusion is that there is a previously unrecognized risk from conducting familial searches created by the possibility that a more distant relative whose profile is in a database will be incorrectly identified as a first-degree relative of the person who left the crime-scene sample.  With the data considered here (13 autosomal loci and 17-locus Y-haplotypes), even with other decision procedures, distinguishability of first-degree and distant genetic relatives may be limited (Figure~\ref{fig:LRdists}).  This is especially troubling when contemplating the possibility that familial searches may be conducted in the national database, which contains over ten million profiles.  Widening the geographic scope of a search is likely to result in more of the source's distant relatives having a presence in the database.

 To be clear, our concerns arise only with respect to inadvertent erroneous identification of distant relatives as first-degree leads.  Familial searches are ineffective if secondary relatives are intentionally sought. Indeed, the Myers protocol targets first-degree relationships only because actively seeking more remote connections ordinarily returns too many leads to investigate.  Yet familial searches also cannot be configured to assure that only first-degree relatives of the crime scene sample source are identified as leads. As our and other research has shown, the tailored approach of the Myers {\it et al.} protocol has the advantage of returning few spurious leads -- if a lead is generated, it is almost certainly a relative of the crime scene sample source.  Our findings, however, suggest that the closeness of the lead to the source is an open question.  Significantly, our research does not reveal the percentage of cases in which a lead returned will be a distant relative, as opposed to a first degree relative. Such an estimation requires a different set of simulations including complex demographic estimates.  

 In our simulations, we set the coancestry coefficient $\theta=.01$, which aligns with the less conservative parameter value suggested for direct identification \cite{NRC1996}.  The currently implemented familial searching methodology in California assumes $\theta=0.0$.  This discrepancy contributes to elevated rates of positive identifications observed between both unrelated individuals and distant Y-sharing relatives.  In addition, our simulation parameter value $\theta=.01$ may be an underestimate for some population samples \cite{NRC1996}.  For these cases, we have underestimated the amount of coincidental relatedness, and thus, estimated power and false positive rate.  This is particularly relevant for some population samples with higher $\theta$ including some Native American groups.  

 In our analysis, we estimate the Y-haplotype frequency upper $95\%$ confidence limit asymptotically, rather than exactly, as indicated in the Myers {\it et al.} method.  This estimate may be sufficient, but has greater error than the exact confidence limit.  For very low Y-haplotype frequencies, the asymptotic estimate may be lower than the true confidence limit, which would lead to inflated (anti-conservative) $LR_Y$.  A study of the affect of different confidence limit estimates on final outcomes would inform method choice.  

 In this study, we have considered only complete genotypes with no errors or allelic dropout.  It is not clear how allelic dropout would affect familial searching results, but this must be explored before considering extention to low-template samples.  

 The probabilities we estimated with our simulations are necessarily approximate. Autosomal allele and Y-haplotype frequencies for various population samples are poorly known because publicly available databases are of limited size and are unavailable for many population groups.  Nevertheless, the groups for which we have data include African Americans, who have relatively high genetic diversity at the considered loci, and Native Americans, who have relatively low diversity, which suggests that our results are applicable to other populations for which data are unavailable.  

 A difference between our analysis and the implemented Myers {\it et al.} method is the one or two-stage design.  In the Myers {\it et al.} method, first an analysis is performed using only autosomal data and the top 168 matches are genotyped for Y-haplotype and the cumulative statistic $X$ is computed only for these samples \cite{Myers2011}.  In our analysis we simply computed the cumulative $X$ for all samples considered.  An additional study of positive identification of distant relatives using the two-stage method in the context of a realistic database would provide more realistic rate estimates, however this sort of analysis is hindered by lack of access to forensic databases \cite{Krane2009}.  Such a study is unlikely to show substantially different results than those presented here since the pairs of individuals we positively identify as first degree relatives are likely to appear related and rank above the 168 person threshold.  

 We also note that in this analysis we only consider the familial searching method of Myers {\it et al.}.  To our knowledge, at the writing of this manuscript, the Myers {\it et al.} method is the only explicit protocol available and the current standard in the field \cite{OConnor2011, GJISI2011}.  Although the absolute rates of identification will change according to the method used, when considering LR-based approaches, which have been shown to be more effective than allele-sharing methods \cite{Balding2012}, the trends we observed across population samples and close and distant relatives will hold.

\subsection*{Implications of spurious identification of distant relatives}
 Our findings confirm that familial searches carried out according to the Myers protocol do a good job of locating a relative if one is in the database.  They also affirm that a search is unlikely to return a false lead -- in other words, a match that appeared to be related to the crime scene source, but in fact was not.  However, we have shown that if there is a more distant relative in the database, that person may have up to a $42\%$ chance of being returned as a lead and erroneously labeled as a first degree relative of the crime scene source (Table \ref{tab:sibIDrates}).  

 The possibility that the lead is a more remote relative of the source might not be a concern if investigators could easily ascertain what kind of lead they had been given. But the Myers protocol can do no more than alert investigators that the source may be a relative of the individual in the database; it does not tell investigators which relative or the closeness or kind of relation.  In any case, once a search returns a lead, law enforcement must undertake further investigation to locate the actual source. It is the scope and impact of the follow-up investigation that, in light of our results, may be troubling.

 Before our research identified the possibility that a familial search might identify distant relatives and erroneously label them as first degree relatives of the source, it may be that law enforcement simply assumed that all leads were to a first degree relative, because that is what the search is structured to find.  Accordingly, if further investigation did not identify a source from among the lead's first degree relatives, then officers likely assumed that the problem was the lead, rather than the depth of their investigation. In light of our results, however, law enforcement may now recognize that a lead that fails to reveal a source among first degree relatives may still be a good lead, it is only that the investigation must extend to more remote branches of the family tree. 

 To illustrate, suppose that law enforcement conducts a familial search to find a burglar. Following the Myers protocol, the search returns a lead to the profile of K, a known offender in the database.  Conventional wisdom holds that the burglar is likely a brother or the father of K, and so law enforcement officers initiate their investigation accordingly.  They ascertain the identify of K's father and any brothers, and check their ages and criminal records.  They determine whether the father or brothers were in the area of the burglary at the time it occurred, used cell phones or credit cards around that area, or otherwise engaged in suspicious behavior.  Ultimately, they might surreptitiously attempt to obtain DNA samples for testing from members of K's immediate family -- say by posing as restaurant personnel or collecting up a half-eaten lunch.  In some number of cases, one of those immediate family members will match, and the burglar will be found.  

 But if no match is made, then investigators aware of our research may conclude three things:  that the familial search was almost certainly effective, that the probability that the lead was a bad lead is low, and that leads that do not initially pan out are likely to have faltered only because the source is a more distant relative than investigators presumed.  In other words, the source is not a brother or father, but instead is a cousin, second cousin, uncle, half-sibling, or even half-cousin.  At that point the officers have two choices.  They may limit themselves to the follow-up they have already conducted with the first degree relatives and simply stop their investigation or, more likely, they may simply widen the scope of their investigation, and start pursuing all second-degree relatives of the lead.

 Our research thus suggests two unanticipated likely outcomes of familial search policies.  First, investigations may wrongly target the immediate families of known offenders, because officers mistakenly believe that their lead is a first-degree relative. Second, investigations may ultimately probe far more deeply than initially imagined, because once officers are convinced that the source cannot be found among first degree relatives, they will widen their net of investigation to include more distant relations. Both of these consequences exacerbate the numerous ethical problems presented by familial searching.

 First, familial searches will affect a greater number of persons.  There is no way for investigators to know from the start that a lead is a distant, rather than immediate, relative of the source.  Thus suspicion may no longer be restricted to a father and small number of siblings -- one of whom is likely to be the crime scene sample source -- but instead will fall upon innocent immediate family members and a much larger number of second-degree relatives. The greater the number of persons involved, and the less likely that one of them is in fact the perpetrator, the more such investigations may begin to feel like a fishing expedition rather than a reasonable search.  This is particularly true given that any investigated family member is, by design, a member of the family whose DNA is not already in the database as a result of wrongdoing.  

 Second, follow-up investigations may prove more intrusive and yet less effective.  Identification of more distant relatives requires more complicated investigation than does determining a lead's immediate family members.  For instance, the known offender will likely have provided information about immediate relatives in the course of the criminal case that is readily available, such as in a bail report, corrections dossier, or probation file.  But such sources are much less likely to contain information about secondary relatives, and thus simply composing the list of potential suspects could require more aggressive investigation.  Moreover, the difficulty in accurately mapping more distant familial relations might lower the already low success rate of familial searches.  Although a lead may in fact be a relative, it may simply be too difficult to locate the actual source if that person is a half-cousin or other distant relation.

 Third, widening the pool intensifies the threat that familial searching poses to our understandings of families as constructions of social, not biological, realities.  A person may have hundreds of ``cousins'' but only a handful of biological cousins.  Investigators may either ignore the difference and unnecessarily investigate those non-biological relations, or else engage in potentially intrusive questioning or activity (such as DNA sampling) to differentiate between proffered and actual relations.  Probing secondary biological relationships might also dredge up painful family experiences of death, unknown biological ties, or previous partners.  And, to the extent that some advocates of familial searching have justified the practice on grounds akin to ``crime runs in families,'' such arguments may be less defensible when more remote connections are involved.  

 Finally, to the extent that our findings suggest that familial searches may in fact necessitate investigating a greater number of people with a greater degree of intrusiveness, that consequence is particularly troubling in that it will be specially visited on certain racial groups.  It has been well documented that familial searching is apt to disproportionately affect African American families, due to the greater representation of those groups in DNA databases and the high rate of intra-racial procreation.  Limiting investigations to the immediate family members of known offenders at least minimizes the intrusion on innocent relatives within those racial groups.  But if more distant relations are included, the web of potential ``genetic suspects'' becomes still broader, and may effectively encompass entire communities.  It takes only one member of a large and varied family tree to render every father, brother, half-brother, cousin, half-cousin, uncle, nephew and so on vulnerable to scrutiny and surreptitious sampling by law enforcement officers.

 Of course, it is always possible to limit, for practical or ethical reasons, the range of permissible follow-up investigation to first degree relatives in familial search cases as a matter of policy.  Such an approach might be sensible from a practical perspective in light of the difficulty in identifying and investigating more remote relatives, and the heightened ethical concerns.  It would also ensure that any spurious leads -- of which, granted, there are expected to be few -- would not first generate highly invasive and costly investigations.  Whatever the case, our research suggests that as states and localities debate the virtues of familial searching and craft policies to govern law enforcement, it would be wise to consider terms delimiting the scope of potential follow-up investigation with regard to degree of relatedness.

\section*{Acknowledgments}
 We are immensely grateful to the individuals whose DNA samples were used in this study, without which none of this work would be possible.  We thank Kirk Lohmueller for his valuable discussions on these topics.  

\section*{Supporting Information Legends}
File S1, Tables S1-5

\newpage
\bibliography{refs}

\newpage
\section*{Figure Legends}

\begin{figure}[!ht] 
  \begin{center}
    \caption{\bf{Distributions  of the test statistic $X$, defined in \eqref{eq:X}, for sibling test for individuals who are siblings (solid red), parent-offsprings (solid black), half-sibs (dashed black), cousins (dashdot black), and second cousins (dotted black).  The population sample individuals are sampled from is along the top and the assumed pop sample is along the side.}}
    \label{fig:LRdists}
  \end{center}
\end{figure}  

\begin{figure}[!ht] 
  \begin{center}
    \caption{\bf{Positive identification rates across different true relationships of individuals simulated from different sample populations Vietnamese American (red circles), African American (orange triangles), European American (purple pluses), Latino American (blue exes), and Native American (green diamonds); left plot is for sibling test, right for parent test}}
    \label{fig:IDrates}
  \end{center}
\end{figure}  

\newpage
\section*{Tables}

\begin{table}[!ht]
    \caption{\bf{False positive parent-offspring identification rates between pairs of unrelated individuals simulated from all pairs of population samples.}}
    \label{tab:unrelpoFPRs}
\centering
	\begin{tabular}{r|rrrrr} \hline
      & Vietnamese & African & European & Latino  & Native \\
      &  American  &  American &  American &  American &  American\\
      \hline
      Vietnamese American & $8.2\e{-7}$ & $5.0\e{-9}$ & $1.5\e{-8}$ & $<5.0\e{-9}$ & $<5.0\e{-9}$ \\
      African American & & $6.5\e{-8}$ & $<5.0\e{-9}$ & $<5.0\e{-9}$ & $<5.0\e{-9}$ \\
      European American & & & $1.5\e{-7}$ & $1.5\e{-8}$ & $1.0\e{-8}$ \\
      Latino American & & & & $1.3\e{-7}$ & $5.0\e{-9}$ \\
      Native American & & & & & $<5.0\e{-9}$ \\ \hline
    \end{tabular}
\end{table}

\begin{table}[!ht]
    \caption{\bf{False positive sibling identification rates between pairs of unrelated individuals simulated from all pairs of population samples. }}
    \label{tab:unrelsibFPRs}
\centering
	\begin{tabular}{r|rrrrr} \hline
      & Vietnamese & African & European & Latino  & Native \\
      &  American  &  American &  American &  American &  American\\
      \hline
      Vietnamese American & $1.1\e{-5}$ & $<5.0\e{-9}$ & $1.0\e{-8}$ & $5.0\e{-9}$ & $<5.0\e{-9}$ \\
      African American & & $1.7\e{-7}$ & $<5.0\e{-9}$ & $5.0\e{-9}$ & $1.0\e{-8}$\\
      European American & & & $1.7\e{-7}$ & $1.5\e{-8}$ & $4.0\e{-8}$ \\
      Latino American & & & & $4.3\e{-7}$ & $2.0\e{-8}$ \\
      Native American & & & & & $<5.0\e{-9}$ \\ \hline
    \end{tabular}
\end{table}

\begin{table}[!ht]
    \caption{\bf{Parent-offspring test identification rates for different Y-sharing relatives and population samples}}
    \label{tab:poIDrates}
\centering
	\begin{tabular}{r|rrrrr} \hline
      & Vietnamese & African & European & Latino  & Native \\
      &  American  &  American &  American &  American &  American\\
      \hline
      parent-offspring & 0.997458 & 0.989336 & 0.987365 & 0.988130 & 0.998809 \\
      sibling & 0.263659 & 0.244048 & 0.255853 & 0.248439 & 0.348373 \\
      half-sib & 0.056135 & 0.045746 & 0.050528 & 0.047072 & 0.105056 \\
      cousin & 0.009311 & 0.006451 & 0.007765 & 0.007039 & 0.027139 \\
      half-cousin & 0.003337 & 0.002019 & 0.002506 & 0.002095 & 0.012716 \\
      second cousin & 0.001971 & 0.000997 & 0.001423 & 0.001145 & 0.008580 \\  \hline
    \end{tabular}
\end{table}

\begin{table}[!ht]
    \caption{\bf{Sibling test identification rates for different Y-sharing relatives and population samples}}
    \label{tab:sibIDrates}
\centering
	\begin{tabular}{r|rrrrr} \hline
      & Vietnamese & African & European & Latino  & Native \\
      &  American  &  American &  American &  American &  American\\
      \hline
      sibling & 0.891566 & 0.819365 & 0.793025 & 0.798273 & 0.925786 \\
      parent-offspring & 0.907037 & 0.813399 & 0.767383 & 0.777360 & 0.942995 \\
      half-sib & 0.303888 & 0.163525 & 0.138446 & 0.140161 & 0.423558 \\
      cousin & 0.099529 & 0.033460 & 0.028376 & 0.027985 & 0.181687 \\
      half-cousin & 0.044457 & 0.010445 & 0.009258 & 0.008978 & 0.100643 \\
      second cousin & 0.027139 & 0.004934 & 0.004582 & 0.004332 & 0.070761 \\ \hline
    \end{tabular}
\end{table}

\end{document}


\pagestyle{empty}

\renewcommand{\thetable}{S\arabic{table}}         
\renewcommand{\thefigure}{S\arabic{figure}}

\begin{center}
{\Large \bf Supporting Information File S1} \\

\vspace{3mm}
{\large 
    The influence of relatives on the efficiency and error rate of familial searching \\
}

\vspace{5mm}

Rori V. Rohlfs$^a$, Erin Murphy$^b$, Yun S. Song$^{c,d}$, Montgomery Slatkin$^a$

\vspace{5mm}
{\small 
$^a${Department of Integrative Biology, University of California, Berkeley, CA 94720, USA}\\
$^b${School of Law, New York University, New York, NY 10012, USA}\\
$^c${Computer Science Division, University of California, Berkeley, CA 94720, USA}\\
$^d${Department of Statistics, University of California, Berkeley, CA 94720, USA}\\
}

\end{center}

\subsection*{Supporting Tables}
\begin{table}[h!]
    \caption{Normalized California census and prison population frequencies.  }
    \label{tab:popFreqs}
  \begin{center}
    \begin{tabular}{r|p{.75in}p{.75in}p{.75in}p{.75in}p{.75in}}   \hline
      & \multicolumn{5}{c}{simulated population} \\
      & Asian & African American & European American & Latino American & Native American \\
      \hline
      Census population & .133 & .0633 & .410 & .384 & .0102 \\
      Prison population & .00626 & .303 & .268 & .413 & .00940 \\  \hline
    \end{tabular}
  \end{center}
\end{table}
Note: We make use of these normative racial categorical constructs to estimate relevant population-specific identification rates.  However, it's clear that membership in these categories is not mutually exclusive.  By nature, assumptions must be made in the collection and tabulation of this sort of data.  Further, as with all collected social data, some groups may be under or over-represented in data collection, and categorical results are subject to biases of reporting method (self-identified, inferred, etc.) (Spade and Rohlfs, {\it in review}).  

\begin{table}[h!]
    \caption{Projected percent of false parent-offspring leads which involve at least one individual from each population sample}
    \label{tab:poFPRsbypop}
  \begin{center}
    \begin{tabular}{r|p{.55in}|p{.7in}p{.7in}p{.7in}p{.7in}p{.7in}}  \hline
      & & \multicolumn{5}{c}{simulated population} \\
      & total FPR & Vietnamese American & African American & European American & Latino American & Native American \\
      \hline
      Census population & 5.73e-08 & 28.3 & 0.560 & 38.5 & 38.59 & 0.114 \\
      Prison population & 3.84e-08 & 0.182 & 16.3 & 25.3 & 62.9 & 0.122 \\  \hline
    \end{tabular}
  \end{center}
\end{table}

\begin{table}[h!]
    \caption{Estimated percent of false sibling identifications which involve at least one individual from each population sample}
    \label{tab:sibFPRsbypop}
  \begin{center}
    \begin{tabular}{r|p{.55in}|p{.7in}p{.7in}p{.7in}p{.7in}p{.7in}} \hline
      & & \multicolumn{5}{c}{simulated population} \\
      & total FPR & Vietnamese American & African American & European American & Latino American & Native American \\
      \hline
      Census population & 3.01e-07 & 67.9 & 0.266 & 10.6 & 22.3 & 0.0849 \\
      Prison population & 1.06e-07 & 0.459 & 15.2 & 13.5 & 73.1 & 0.199 \\ \hline
    \end{tabular}
  \end{center}
\end{table}

\begin{table}[h!]
    \caption{Parent-offspring test identification rates for different Y-sharing relatives and population samples where the set of reference Y-haplotypes was down-sampled to 103 haplotypes per population sample.  The similarity of these results to those with the full set of Y-haplotypes indicates that the varying reference sample sizes are not driving population-based differences in identification rates.  }
    \label{tab:subpoIDrates}
  \begin{center}
    \begin{tabular}{r|p{.75in}p{.75in}p{.75in}p{.75in}p{.75in}} \hline
      & Vietnamese American & African American & European American & Latino American & Native American \\
      \hline
      parent-offspring & 0.971109 & 0.917481 & 0.900600 & 0.904750 & 0.981429 \\
      sibling & 0.260568 & 0.238175 & 0.247218 & 0.240125 & 0.346432 \\
      half-sib & 0.052082 & 0.037330 & 0.039416 & 0.037430 & 0.099234 \\
      cousin & 0.008184 & 0.004503 & 0.005204 & 0.004543 & 0.024529 \\
      half-cousin & 0.002859 & 0.001183 & 0.001447 & 0.001275 & 0.010976 \\
      second cousin & 0.001570 & 0.000522 & 0.000741 & 0.000616 & 0.007245 \\ \hline
    \end{tabular}
  \end{center}
\end{table}

\begin{table}[h!]
    \caption{Sibling test identification rates for different Y-sharing relatives and population samples where the set of reference Y-haplotypes was down-sampled to 103 haplotypes per population sample.  The similarity of these results to those with the full set of Y-haplotypes indicates that the varying reference sample sizes are not driving population-based differences in identification rates.  }
    \label{tab:subsibIDrates}
  \begin{center}
    \begin{tabular}{r|p{.75in}p{.75in}p{.75in}p{.75in}p{.75in}}  \hline
      & Vietnamese American & African American & European American & Latino American & Native American \\
      \hline
      sibling & 0.786012 & 0.677311 & 0.636255 & 0.645179 & 0.831449 \\
      parent-offspring & 0.750433 & 0.595312 & 0.521906 & 0.536782 & 0.815689 \\
      half-sib & 0.166490 & 0.070316 & 0.055062 & 0.056729 & 0.240527 \\
      cousin & 0.043701 & 0.010769 & 0.008384 & 0.008458 & 0.080298 \\
      half-cousin & 0.016970 & 0.002642 & 0.002165 & 0.002175 & 0.038323 \\
      second cousin & 0.009340 & 0.001130 & 0.000979 & 0.000928 & 0.024955 \\ \hline
    \end{tabular}
  \end{center}
\end{table}